\documentclass[reprint, superscriptaddress, secnumarabic, amssymb, nobibnotes, aps, prl]{revtex4-1}

\usepackage{graphicx}
\usepackage{epstopdf}
\usepackage[T1]{fontenc}
\usepackage[latin9]{inputenc}
\usepackage{amsbsy}
\usepackage{gensymb}
\usepackage[T1]{fontenc}
\usepackage[latin9]{inputenc}
\usepackage{amsmath}
\usepackage{amssymb}
\usepackage{bbm}
\usepackage{braket}
\usepackage{xcolor}
\allowdisplaybreaks
\usepackage{graphicx}
\usepackage[colorlinks=true]{hyperref}  
\hypersetup{
    bookmarks=true,         
    unicode=false,          
    pdftoolbar=true,        
    pdfmenubar=true,        
    pdffitwindow=false,     
    pdfstartview={FitH},    
    pdftitle={unconventional superconducting properties in noncentrosymmetric superconductor Re5.5Ta},    
    pdfauthor={Arushi, D. Singh, P. K. Biswas, A. D. Hillier, R. P. Singh},     
    pdfsubject={},   
    pdfcreator={},   
    pdfproducer={}, 
    pdfkeywords={} {} {}, 
    pdfnewwindow=true,      
    colorlinks=true,       
    linkcolor=blue, 
    citecolor=blue,        
    filecolor=magenta,      
    urlcolor=blue           
} 
\usepackage[normalem]{ulem}

\newcommand{\equref}[1]{Eq.~(\ref{#1})}

\newcommand{\figref}[1]{Fig.~\ref{#1}}

\begin{document}

\title{\textrm{Probing the superconducting ground state of noncentrosymmetric high entropy alloys using muon-spin rotation and relaxation}}
\author{Kapil~Motla}
\affiliation{Department of Physics, Indian Institute of Science Education and Research Bhopal, Bhopal, 462066, India}
\author{Arushi}
\affiliation{Department of Physics, Indian Institute of Science Education and Research Bhopal, Bhopal, 462066, India}
\author{P. K. Meena}
\affiliation{Department of Physics, Indian Institute of Science Education and Research Bhopal, Bhopal, 462066, India}
\author{D.~Singh}
\affiliation{ISIS Facility, STFC Rutherford Appleton Laboratory, Harwell Science and Innovation Campus, Oxfordshire, OX11 0QX, UK}
\author{P.~K.~Biswas}
\affiliation{ISIS Facility, STFC Rutherford Appleton Laboratory, Harwell Science and Innovation Campus, Oxfordshire, OX11 0QX, UK}
\author{A.~D.~Hillier}
\affiliation{ISIS Facility, STFC Rutherford Appleton Laboratory, Harwell Science and Innovation Campus, Oxfordshire, OX11 0QX, UK}
\author{R.~P.~Singh}
\email[]{rpsingh@iiserb.ac.in}
\affiliation{Department of Physics, Indian Institute of Science Education and Research Bhopal, Bhopal, 462066, India}

\date{\today}
\begin{abstract}
\begin{flushleft}
\end{flushleft}
Recently, high entropy alloys (HEAs) have emerged as a new platform for discovering superconducting materials and offer avenues to explore exotic superconductivity. The highly disordered nature of HEA suggests regular phonon required for BCS superconductivity may be unlikely to occur. Therefore understanding the microscopic properties of these superconducting HEA is important. We report the first detailed characterization of the superconducting properties of the noncentrosymmetric ($\alpha$-Mn structure) HEA {(HfNb)}$_{0.10}${(MoReRu)}$_{0.90}$, and {(ZrNb)}$_{0.10}${(MoReRu)}$_{0.90}$ by using magnetization, specific heat, AC transport, and muon-spin relaxation/rotation ($\mu$SR). Despite the disordered nature, low temperature specific heat and transverse-field muon spin rotation measurements suggest nodeless isotropic superconducting gap and Zero-field $\mu$SR measurements confirm that time reversal symmetry is preserved in the superconducting ground state.
\end{abstract}

\maketitle

\section{Introduction}
High entropy alloys (HEAs) are a new class of materials with tunable physical and superior mechanical properties compared to conventional binary and ternary alloys. These materials are getting widespread attention from many different scientific areas, material science, theoretical and experimental condensed matter physics \cite{thermal stability, high strength,ductility1,ductility2, thermoelectric,High sat.mag}. HEA are multi-component alloys that contain five or more elements in near equimolar ratios \cite{ nano, over conventional, four element4+conf entropy2, 4 elements2}. The Gibbs free energy decreases at high temperature and plays a vital role in crystallizing HEA in different crystallographic structures \cite{prx, high strength, Alloy design}.
Recently, HEA has emerged as a new class of disordered alloy superconductors, having a high superconducting transition temperature and critical field. Also, it shows retention of superconductivity at very high pressure  \cite{hp}. Superconductivity was first reported in the high entropy alloy, Ta$_{34}$Nb$_{33}$Hf$_{8}$Zr$_{14}$Ti$_{11}$ \cite{prl}, since then it has been observed in a few others high entropy alloys \cite{V hea, nmrrp, hexa hea, cava, Mizu, rcava, kita}. 
To date, most of the research in HEA superconductors has been focused on discovering new HEA superconducting materials that crystallises in various structures and enhancing the superconducting transition temperature. In contrast, the superconducting pairing mechanism is largely unexplored, mainly due to the HEA multi-component and a high disorder nature. It is difficult to calculate the electronic structure and understand the lattice vibration, which is usually essential for understanding the superconducting pairing mechanism. A comparative study with binary alloy superconductors with HEA, which have the same crystal structure and a large disorder, can provide more insight into the superconducting properties of HEA. A topical example of a binary alloy is the Re-based superconductors. These materials have a noncentrosymmetric $\alpha$-Mn crystal structure and have been studied extensively due to the presence of time-reversal symmetry (TRS) breaking \cite{Bauer,Usc1,Usc2,Usc3,skgosh}. However, the exact superconducting pairing mechanism is still not fully understood. Structural similarity, disorder, and the multi-component nature of Re-based noncentrosymmetric (NCS) HEA, may help understand the superconducting pairing mechanism of noncentrosymmetric superconducting compounds and HEA itself, which is still elusive.
In this paper, we have performed a comprehensive study superconducting ground state, using magnetization, heat capacity, and resistivity, together with muon-spin spectroscopy on a new NCS $\alpha$-$Mn$ HEA {(HfNb)}$_{0.10}${(MoReRu)}$_{0.90}$ having a $T_{C}$ = 5.9(1) $K$ and another $\alpha$-$Mn$ NCS HEA {(ZrNb)}$_{0.10}${(MoReRu)}$_{0.90}$ reported by Stolze et al. \cite{cava}) having a $T_{C}$ 5.8(1)~$K$. \\

\section{Experimental Details}
Polycrystalline samples of {(HfNb)}$_{0.10}${(MoReRu)}$_{0.90}$ and {(ZrNb)}$_{0.10}${(MoReRu)}$_{0.90}$ HEA were prepared by arc-melting stoichiometric quantities of high-purity elements (5N) under argon (5N) atmosphere. The resulting ingots were flipped and melted several times to enhance the homogeneity. In both HEAs, the weight loss was negligible (< 0.1\%) after melting. Phase purity and crystal structure of the samples were confirmed by X-ray diffraction (XRD) at room temperature on a PANalytical diffractometer equipped with $CuK_{\alpha}$ radiation ({$\lambda$} = 1.54056~\AA).

Temperature and field dependent magnetization, heat capacity and transport measurements were performed using a Quantum Design, MPMS-3 and PPMS. The $\mu SR$ measurements in zero-field (ZF) and transverse-field (TF) conditions were carried out using the MUSR spectrometer at the ISIS Neutron and Muon Facility, STFC Rutherford Appleton Laboratory, United Kingdom. A full description of the $\mu$SR technique may be found in Ref. \cite{aidy}.

\section{Results}
\subsection{Structural characterization}

\begin{figure} 
\centering
\includegraphics[width=1.0\columnwidth, origin=b]{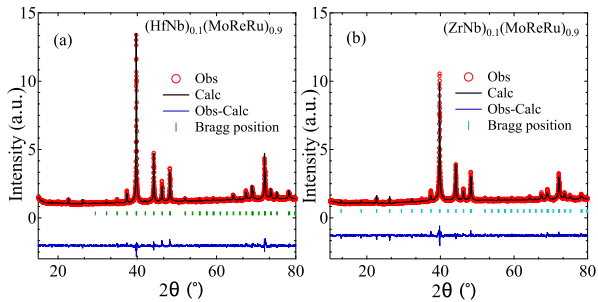}
\caption{Color online: The X-ray diffraction pattern from {(HfNb)}$_{0.10}${(MoReRu)}$_{0.90}$ (left) and {(ZrNb)}$_{0.10}${(MoReRu)}$_{0.90}$ (right) collected at room temperature using $CuK_{\alpha}$ radiation ({$\lambda$} = 1.54056~\AA). The black line is the Le Bail fitting and the green dashes are the expected locations for the diffraction Bragg peaks.}
\label{Fig1}
\end{figure}

X-ray diffraction spectra from both HEA were collected at ambient conditions and Le Bail fitting using the Fullprof software \cite{Fullprof} as shown in \figref{Fig1}. This shows that both alloys crystallize in a cubic non-centrosymmetric $\alpha$-Mn (space group I$\bar{4}3$\textit{m}) crystal structure and have unit cell parameters: $a$ = 9.6170(2) \(\text{\AA}\) and 9.6180(2) \(\text{\AA}\) for {(HfNb)}$_{0.10}${(MoReRu)}$_{0.90}$ and {(ZrNb)}$_{0.10}${(MoReRu)}$_{0.90}$ respectively. The refined cell parameters of {(ZrNb)}$_{0.10}${(MoReRu)}$_{0.90}$ are in good agreement with the published data \cite{cava}. Due to large number of different atoms/sites and tendency to form solid solution makes difficult to determine the occupancies of atomic sites unambiguously (See supplementary information \cite {supply}).

\subsection{Normal and superconducting state properties}
\subsubsection{Electrical Resistivity}
\begin{figure} 
\includegraphics[width=1.0\columnwidth, origin=b]{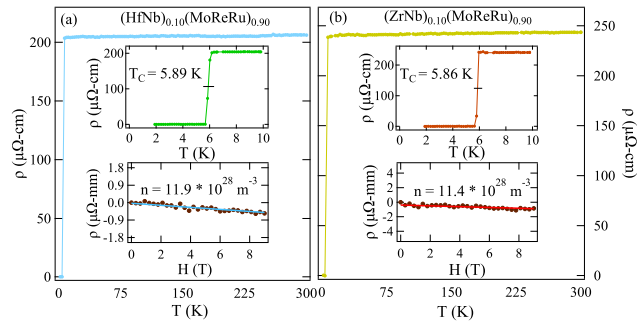}
\caption{ Temperature dependence of the resistivity for a) {(HfNb)}$_{0.10}${(MoReRu)}$_{0.90}$, b) {(ZrNb)}$_{0.10}${(MoReRu)}$_{0.90}$ in zero field. The top of both Insets is showing the superconducting transition and the bottom of both insets is showing the field-dependent hall resistivity at 10 $K$.}
\label{Fig2:res}
\end{figure}

Temperature dependence of the resistivity, $\rho (T)$, for both samples were performed in zero field from  1.9 $K$ to 300 $K$. The results for {(HfNb)}$_{0.10}${(MoReRu)}$_{0.90}$ and {(ZrNb)}$_{0.10}${(MoReRu)}$_{0.90}$ are shown in \figref{Fig2:res}(a) and \ref{Fig2:res}(b) respectively. Low temperature data is shown in the top inset of \figref{Fig2:res}(a) and \figref{Fig2:res}(b) clearly presents a very sharp  drop in resistivity at $T_{C}^{mid}$ = 5.9(1) $K$ and 5.8(1) $K$ for {(HfNb)}$_{0.10}${(MoReRu)}$_{0.90}$ and {(ZrNb)}$_{0.10}${(MoReRu)}$_{0.90}$, respectively. Resistivity increases leisurely with temperature showing poor metallic behavior. The residual resistivity ratio (RRR) for both the samples found to be ~1.2. The small value of RRR for both HEAs indicates a high degree of disorder and these values are comparable to reported RRR ratio for HEAs and $\alpha$-Mn binary alloys \cite{Usc1,Usc2,Usc3,RRR,cava}. Hall measurement was also performed to calculate the carrier concentration and the type of charge carriers. Lower in inset in \figref{Fig2:res}(a) and \ref{Fig2:res}(b) shows the field dependence of hall resistivity ($\rho$) measured at T = 10 K for both HEA's. $\rho$(H) is well described by a straight line fit and carrier concentration yields $n = 11.9(4)\times10^{-28} m^{-3}$ and 11.4(8)$\times10^{-28} m^{-3}$ respectively for {(HfNb)}$_{0.10}${(MoReRu)}$_{0.90}$ and {(ZrNb)}$_{0.10}${(MoReRu)}$_{0.90}$.

\begin{figure*} 
\includegraphics[width=2.0\columnwidth, origin=b]{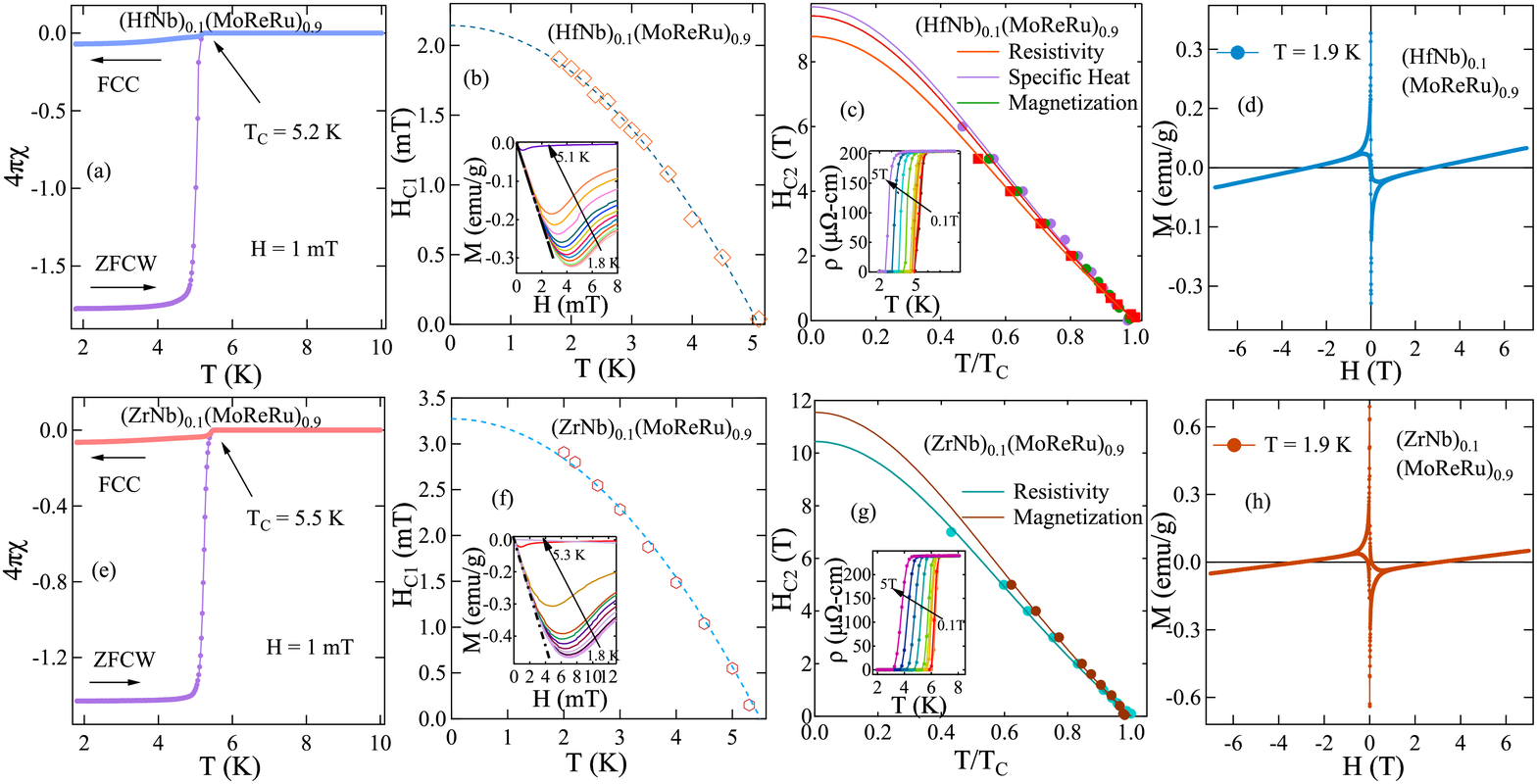}
\caption{ (a) and (e) show the temperature dependence of magnetic moment in 1.0 $mT$ in FCC and ZFCW mode. (b) and (f) show the temperature dependence of lower critical field. (c) and (g) show the upper critical field estimated using specific heat, resistivity, and magnetization data. The dotted lines is the result of the fit to equation \ref {hc2}. (d) and (h) Magnetic field dependent magnetization (M-H) at 1.9 K.}
\label{Fig3}
\end{figure*}

\subsubsection{Magnetization}

To confirm bulk superconductivity in both HEA, the temperature dependence of the DC magnetization measurements were performed in an applied field of 1.0 $mT$ in zero field cooled warming (ZFCW) and field cooled cooling (FCC) modes. The onset of superconductivity was observed below $T_{C}$ = 5.2(1) $K$ for {(HfNb)}$_{0.10}${(MoReRu)}$_{0.90}$ and 5.5(1) $K$ for {(ZrNb)}$_{0.10}${(MoReRu)}$_{0.90}$ by a sharp decrease in a diamagnetic magnetisation, as shown in Fig. \ref{Fig3}(a) and \ref{Fig3}(e). In order, to estimate the lower critical field $H_{C1}$, we have performed magnetization versus field measurements at a range of  temperatures. The value of the $H_{C1}$ at each temperature is taken as the deviation of the magnetisation from the linearity, as shown in the inset of Fig. \ref{Fig3}(b) and \ref{Fig3}(f) for  {(HfNb)}$_{0.10}${(MoReRu)}$_{0.90}$ and {(ZrNb)}$_{0.10}${(MoReRu)}$_{0.90}$. The lower critical field at absolute zero temperature, $H_{C1}$(0) can be calculated by extrapolating $H_{C1}$(T) using the Ginzburg-Landau expression, which is given as\\
\begin{equation}
H_{C1}(T)=H_{C1}(0)\left(1-t^{2}\right)
\label{Hc1}
\end{equation} 
where $t$ = $T/T_{C}$, and lower critical field $H_{C1}$(0) was estimated as 2.14(1) $mT$ and 3.27(3) $mT$ for {(HfNb)}$_{0.10}${(MoReRu)}$_{0.90}$ and {(ZrNb)}$_{0.10}${(MoReRu)}$_{0.90}$ respectively by fitting the Eq. \ref{Hc1} in the data given in Fig. \ref{Fig3}(b) and \ref{Fig3} (f).

The upper critical field at $T$ = 0 $K$, $H_{C2}(0)$ estimated using Ginzburg-Landau(GL) relation\\
\begin{equation}
H_{C2}(T) = H_{C2}(0)\left(\frac{(1-t^{2})}{(1+t^2)}\right) \label{hc2}
\end{equation}\\
where $t$ = $T/T_{C}$, and the estimated value of $H_{C2(res,mag,hc)}$(0) = 8.7(1) $T$, 9.4(1) $T$, 9.7(1) $T$ for {(HfNb)}$_{0.10}${(MoReRu)}$_{0.90}$ and $H_{C2(Res,mag)}$(0) = 10.4(9) $T$, 11.5(2) $T$ for {(ZrNb)}$_{0.10}${(MoReRu)}$_{0.90}$. The Ginzburg-Landau coherence length (the length between cooper pair) can be estimated with the help of $H_{C2}$(0) by using the expression
\begin{equation}
H_{C2}(0) = \frac{\Phi_{0}}{2\pi\xi_{GL}^{2}}
\label{eqn3:up}
\end{equation}

where $\Phi_0$ is the flux quantum ($\Phi_0$ = 2.07$\times$10$^{-15}$T-m$^2$). After substituting $H_{C2}$(0) value from magnetization gives $\xi_{GL}$ = 5.92(2) $nm$ for {(HfNb)}$_{0.10}${(MoReRu)}$_{0.90}$ and 5.35(3) $nm$ for {(ZrNb)}$_{0.10}${(MoReRu)}$_{0.90}$. The calculated values of $H_{C1}$(0) and $\xi _{GL}$(0) were used to evaluate magnetic penetration depth $\lambda_{GL}$(0) (the value of distance at which magnetic field become 1/$e$ times of external applied magnetic field) for both the samples with the help of relation\\
\begin{equation}
H_{C1}(0) = \frac{\Phi_{0}}{4\pi\lambda_{GL}^2(0)}\left(\mathrm{ln}\frac{\lambda_{GL}(0)}{\xi_{GL}(0)}+0.12\right)   
\label{eqn6:ld}
\end{equation} 
and were obtained as 609(2) $nm$ and and 487(3) $nm$ for {(HfNb)}$_{0.10}${(MoReRu)}$_{0.90}$ and {(ZrNb)}$_{0.10}${(MoReRu)}$_{0.90}$ respectively. Ginzburg-Landau ratio is given by $\frac{\lambda_{GL}(0)}{\xi_{GL}(0)}$, $\kappa_{GL}$ $>$ ${\frac{1}{\sqrt{2}}}$ for {(HfNb)}$_{0.10}${(MoReRu)}$_{0.90}$ and {(ZrNb)}$_{0.10}${(MoReRu)}$_{0.90}$ respectively. This confirms that both of these HEA are indeed strong type II superconductors.

In a type-II superconductor, Cooper pair breaking due to the applied magnetic field is attributed to two types of mechanisms: orbital limiting field and Pauli paramagnetic limiting field effect \cite{breaking cooper1,breaking cooper2}. In the orbital pair breaking, the induced kinetic energy of a Cooper pair by an external field exceeds to the Cooper pair condensation energy. Whereas in Pauli limiting, the applied magnetic field aligns one of the Cooper pair spin moments in the direction of its field, thereby breaking the pairing. Orbital limiting field, $H_{C2}^{orbital}$(0) is given by the Werthamar-Helfand-Hohenberg (WHH) expression \\
\begin{equation}
H_{C2}^{orbital}(0) = -\alpha T_{C}\left.\frac{dH_{C2}(T)}{dT}\right|_{T=T_{C}}
\label{eqn4:whh}
\end{equation}

 where $\alpha$ is purity factor and 0.693 value define for dirty limit superconductors (see discussion section). The initial slope -${\frac{dH_{C2}}{dT}}$ at $T$ = T$_C$ were estimated 2.24(5) $T/K$, and 2.6(2) $T/K$ for {(HfNb)}$_{0.10}${(MoReRu)}$_{0.90}$, and {(ZrNb)}$_{0.10}${(MoReRu)}$_{0.90}$ respectively, and gives the orbital limiting upper critical field $H_{C2}^{orb}$(0) as 8.0(2) $T$ for {(HfNb)}$_{0.10}${(MoReRu)}$_{0.90}$ and 9.9(5) $T$ for {(ZrNb)}$_{0.10}${(MoReRu)}$_{0.90}$. The Pauli paramagnetic limit is given by $H_{C2}^{P}$ = 1.84 $T_{C}$ within the BCS theory. Substituting the values of $T_{C}$, we have determined  $H_{C2}^{P}$ = 9.56(2) and 10.12(2) $T$ for {(HfNb)}$_{0.10}${(MoReRu)}$_{0.90}$, and {(ZrNb)}$_{0.10}${(MoReRu)}$_{0.90}$ respectively. The Maki parameter, which is a  measure of the strength of Pauli limiting field and orbital critical field is given by the expression $\alpha_{M} = \sqrt{2}H_{C2}^{orb}(0)/H_{C2}^{p}(0)$
The values obtained for $\alpha_{M}$ are 1.18 for {(HfNb)}$_{0.10}${(MoReRu)}$_{0.90}$ and 1.38 for {(ZrNb)}$_{0.10}${(MoReRu)}$_{0.90}$.

The magnetization hysteresis loops ($\pm$ 9.0 $T$) for both HEAs at 1.9 $K$ is shown in Fig. \ref{Fig3}(d) and \ref{Fig3}(h). A closed-loop form at 2.0 $T$ and 3.5 $T$, denotes the $H_{irr}$ is far below from their upper critical magnetic fields of both HEA samples. These values of $H_{irr}$ is suggesting the depinning of the flux line vortexes. The depinning generally happens due to the thermal fluctuation on the condensation energy of Cooper pair or stress induced by grain boundary/disorder in polycrystalline samples. The strength of thermal fluctuation with respect to condensation energy of charge carriers is described by the $G_{i}$ number \cite{Om prakash}. The calculated value of $G_{i}$ for both HEAs ($\sim$ 10$^{-5}$) and falls between high $T_{C}$  ($\sim$ 10$^{-2}$) and conventional superconductors ($\sim$ 10$^{-8}$), suggesting the defect and grain boundaries can be responsible for depinning \cite{Bi8Si42Al4}.\\ 

\subsubsection{Specific Heat}
\begin{figure} 
\centering
\includegraphics[width=1.0\columnwidth, origin=b]{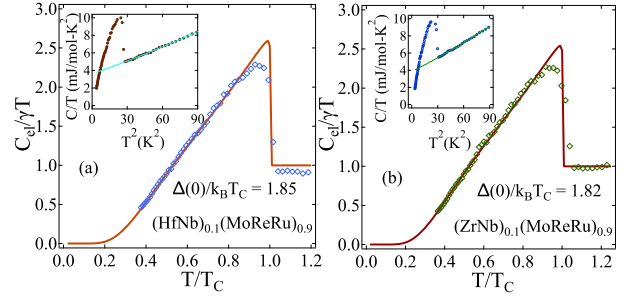}
\caption{Normalised specific heat data $C_{el}$/$\gamma_nT$ fitted with the BCS $s$-wave model shown by the red line for both HEAs (a) {(HfNb)}$_{0.10}${(MoReRu)}$_{0.90}$ and (b) {(ZrNb)}$_{0.10}${(MoReRu)}$_{0.90}$. The inset shows the temperature dependent specific heat data in zero field data plotted as $C/T$ vs $T^2$.}
\label{Fig4}
\end{figure}

Specific heat measurement for both samples were performed between 1.9 - 20 $K$ in zero-field. The observed $T_C$ for both the samples is in agreement with the magnetization and resistivity data. The specific heat data above $T_{C}$ in the normal region was fitted using the equation: $\frac{C}{T}=\gamma_{n}+\beta_{3} T^{2} + \beta_{5}T^{4}$ and is shown in the insets of Fig. \ref{Fig4}. Here $\gamma_n$ is the coefficient for electronic specific heat in the normal state  (Sommerfeld coefficient) and $\beta_3$ and $\beta_5$ are the phononic contribution. The fitting provides the parameters as: $\gamma_n$ = (3.6(1), 3.8(1)) $mJ$-$mol^{-1} K^{-2}$, $\beta_3$ = (0.047(1), 0.052(3)) $mJ$-$mol^{-1}K^{-4}$, and $\beta_5$ = (0.07(1), 0.07(1)) $\mu$$J$-$mol^{-1} K^{-6}$ for {(HfNb)}$_{0.10}${(MoReRu)}$_{0.90}$, {(ZrNb)}$_{0.10}${(MoReRu)}$_{0.90}$ respectively.

The density of state $D_C$(E$_F$) and Debye temperature, $\theta_D$ have been calculated using $\gamma_n$ and $\beta_3$. The obtained value of $D_C$(E$_F$) = (1.53(3), 1.63(4)) $\frac{states}{eV f.u}$ and $\theta_D$ = 346(2) $K$, 335(7) $K$ for for {(HfNb)}$_{0.10}${(MoReRu)}$_{0.90}$, {(ZrNb)}$_{0.10}${(MoReRu)}$_{0.90}$ respectively. Moreover, the strength of the attractive interaction between electron and phonon can be expressed according to the McMillan model \cite{McMillan} as\\
\begin{equation}
\lambda_{e-ph} = \frac{1.04+\mu^{*}\mathrm{ln}(\theta_{D}/1.45T_{C})}{(1-0.62\mu^{*})\mathrm{ln}(\theta_{D}/1.45T_{C})-1.04 }
\label{eqn8:ld}
\end{equation}
\\
here $\mu^{*}$ is the screened Coulomb repulsion parameter which is usually between 0.1-0.15 and for intermetallic superconductors $\sim$ 0.13 \cite{Usc2,Usc3}. Inserting the value of Debye temperature, $\theta_D$ and $T_{C}$, we find the strength between electron and phonon, $\lambda_{e-ph}$ = 0.62(6) and 0.63(8) for {(HfNb)}$_{0.10}${(MoReRu)}$_{0.90}$ and {(ZrNb)}$_{0.10}${(MoReRu)}$_{0.90}$ respectively. This value indicates a moderately coupled superconductivity similar to other Re based noncentrosymmetric superconductors such as Re$_{6}$Hf \cite{Usc2} and Re$_{24}$Ti$_{5}$ \cite{moderately}.

In order to determine the electronic specific heat contribution, we have subtracted the phononic contribution from the total specific heat: $C_{el}$ = $C(T)$ -$\beta_{3}T^3$ - $\beta_{5}T^5$. The normalized electronic specific heat jump, $\frac{\Delta C_el}{\gamma_n T_C}$ = 1.67 for {(HfNb)}$_{0.10}${(MoReRu)}$_{0.90}$ and 1.49 for {(ZrNb)}$_{0.10}${(MoReRu)}$_{0.90}$ which further suggests moderately coupled superconductivity for both these HEA samples. The electronic specific heat data below transition temperature $T_C$ can be best fitted with single-gap BCS expression for normalized entropy, $S$ 
\\
\begin{equation}
\frac{S}{\gamma_{n}T_{C}} = -\frac{6}{\pi^2}\left(\frac{\Delta(0)}{k_{B}T_{C}}\right)\int_{0}^{\infty}[ \textit{f}\ln(f)+(1-f)\ln(1-f)]dy \\
\label{eqn7:s}
\end{equation}
\\
where $\textit{f}$($\xi$) = [exp($\textit{E}$($\xi$)/$k_{B}T$)+1]$^{-1}$ is the Fermi function, $\textit{E}$($\xi$) = $\sqrt{\xi^{2}+\Delta^{2}(t)}$, where $E(\xi $) is the energy of the normal electrons relative to the Fermi energy, $\textit{y}$ = $\xi/\Delta(0)$, $\mathit{t = T/T_{C}}$ and $\Delta(t)$ = $tanh$[1.82(1.018(($\mathit{1/t}$)-1))$^{0.51}$] is the BCS approximation for the temperature dependence of the energy gap. The normalized electronic specific heat below $T_{C}$ is related to the normalized entropy by 
\\
\begin{equation}
\frac{C_{el}}{\gamma_{n}T_{C}} = t\frac{d(S/\gamma_{n}T_{C})}{dt} \\
\label{eqn8:Cel}
\end{equation}
\\
Fig. \ref{Fig4} shows the Eq. \ref{eqn8:Cel} fits to the specific heat data. This provides $\frac{\Delta(0)}{k_{B} T_{C}}$ = 1.85(3) and 1.82(2) for {(HfNb)}$_{0.10}${(MoReRu)}$_{0.90}$, {(ZrNb)}$_{0.10}${(MoReRu)}$_{0.90}$ respectively, both of which are higher than the usual BCS value in the weak coupling limit, again suggesting moderately coupled superconductivity in both HEA.
\\
\begin{figure} 
\centering
\includegraphics[width=1.0\columnwidth, origin=b]{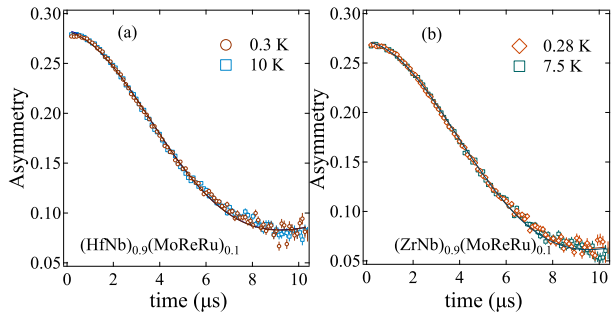}
\caption{Temperature dependent ZF-$\mu$SR asymmetry spectra collected above and below the transition temperature for both the samples. The fitting curve is shown as solid red curve.}
\label{Fig5}
\end{figure}

\subsubsection{Muon spin relaxation and rotation}
The nature of superconducting ground state of both HEAs, {(HfNb)}$_{0.10}${(MoReRu)}$_{0.90}$ and {(ZrNb)}$_{0.10}${(MoReRu)}$_{0.90}$ was further investigated by using muon spin relaxation and rotation measurements. First, we shall discuss the ZF-$\mu$SR measurements, which were carried out above and below $T_{C}$ for both samples. This was to detect any possibility of the presence of time reversal symmetry breaking signal. The absence of any precession signal confirms the absence of local magnetic field associated with long-range ordering, and depolarization of muon spin occurs due to the presence of static randomly oriented nuclear moments. In the absence of magnetic moment, the behavior of time-dependent muon asymmetry spectra is best described by Gaussian Kubo-Toyabe function \cite{Kubo} \\
\begin{equation}
G_{\mathrm{KT}}(t) = \frac{1}{3}+\frac{2}{3}(1-\sigma^{2}_{\mathrm{ZF}}t^{2})\mathrm{exp}\left(\frac{-\sigma^{2}_{\mathrm{ZF}}t^{2}}{2}\right)
\label{eqn17:zf}
\end{equation} 
\\
where $\sigma_{ZF}$ is the relaxation rate of muon-spin due to static, randomly oriented local fields associated with the nuclear moments. The time-dependent asymmetry spectra can be best described by the following function

\begin{equation}
A(t) = A_{1}G_{\mathrm{KT}}(t)\mathrm{exp}(-\Lambda t)+A_{\mathrm{BG}} 
\label{eqn9:tay}
\end{equation}

\begin{figure}
\includegraphics[width=1.0\columnwidth, origin=b]{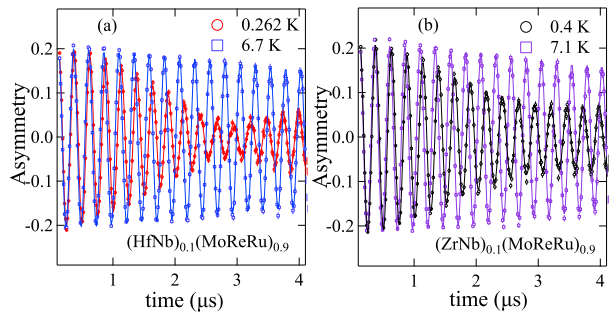}
\caption{Transverse field asymmetry spectra were collected at 30 $mT$ magnetic field above and below the transition temperature of HEAs. The solid line is the fit which  using the a Gaussian modulated oscillatory function.}
\label{Fig6}
\end{figure}
where $A_1$ and $A_{BG}$ are the sample asymmetry and non-decaying constant background signal while $\Lambda$ is an electronic relaxation rate. ZF-$\mu$SR spectra collected both in normal and superconducting state exhibit the identical relaxations can be seen in  overlapping ZF-$\mu$SR spectra (Fig. \ref{Fig5}). It confirms additional ZF-$\mu$SR relaxations below superconducting transition temperature exclude the possibility of time reversal symmetry in superconducting ground state of both the HEA's.

To gain information on the superconducting gap structure, we have performed TF-$\mu$SR measurement where an applied magnetic field of 30 $mT$ was applied above the superconducting transition temperature, and then the sample was cooled to the base temperature 0.26 $K$ of the $He^{3}$ cryostat. The applied magnetic field is greater than $H_{C1}$ but less than $H_{C2}$ in order to generate a flux line lattice. Fig. \ref{Fig6} shows the TF-$\mu$SR asymmetry spectra below and above $T_{C}$ for both HEAs. The fast decay of TF-$\mu$SR spectra below T$_{C}$ with respect to above T$_{C}$ spectra is due to the formation of flux lattice line. The TF-$\mu$SR signal is best fitted with the oscillatory function:

\begin{equation}
\begin{split}
A(t) = \sum_{i=1}^N A_{i}\exp\left(-\frac{1}{2}\sigma_i^2t^2\right)\cos(\gamma_\mu B_it+\phi)\\ + A_{bg}\cos(\gamma_\mu B_{bg}t+\phi)
\label{eqn11:TF1}
\end{split}
\end{equation}

where $B_{i}$ is mean field of the $i_{th}$ component of the Gaussian distribution, $B_{bg}$ is the contribution from the sample holder, $A_{i}$ and $A_{bg}$ are the asymmetry contribution from the sample and sample holder, $\phi$ is the initial phase offset, and $\sigma$ is the Gaussian muon spin depolarization rate. The second moment was used in case of {(HfNb)}$_{0.10}${(MoReRu)}$_{0.90}$ and the first and second moments are given as\\
\begin{equation}
{B} = \sum_{i=1}^2\frac{A_{i}B_{i}}{A_{1}+A_{2}},
\label{eqn12:TF2}
\end{equation}

\begin{equation}
<{\Delta B^2}> = \frac{\sigma^2}{\gamma_{\mu}^2} = \sum_{i=1}^2\frac{A_{i}[(\sigma_{i}/\gamma_{\mu})^2+(B_{i}-{B})^2]}{A_{1}+A_{2}}
\label{eqn13:TF3}
\end{equation}\\
$\sigma$ includes both the temperature-independent depolarization, $\sigma_{N}$, which is coming from the static field arises due to the nuclear magnetic moment and the contribution of the field variation from the flux line lattice and is given as $\sigma^2 = \sigma_{N}^2 + \sigma_{FLL}^2$. As for both the samples, $\xi(0)$/$l$ > 1 (see Table 1), temperature dependent London magnetic penetration depth in dirty limit within London approximation, can be estimated by following expression
\\
\begin{equation}
\frac{\sigma_{FLL}(T)}{\sigma_{FLL}(0)} = \frac{\lambda^{-2}(T)}{\lambda^{-2}(0)} = \frac{\Delta(T)}{\Delta(0)}\mathrm{tanh}\left[\frac{\Delta(T)}{2k_{B}T}\right] 
\label{eqn14:lpd}
\end{equation}
\\
\begin{figure}
\includegraphics[width=1.0\columnwidth, origin=b]{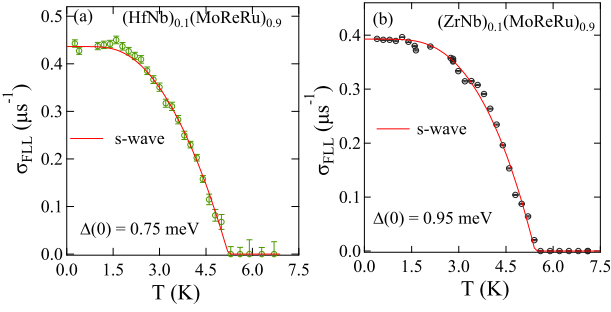}
\caption{TF field muon depolarization data collected at 30 $mT$. The data is well described using an isotropic $s$-wave model for both {(HfNb)}$_{0.10}${(MoReRu)}$_{0.90}$ (left) and {(ZrNb)}$_{0.10}${(MoReRu)}$_{0.90}$ (right).}
\label{Fig7}
\end{figure}

The solid line in Fig. \ref{Fig7} is the fit using \equref{eqn14:lpd} to the muon depolarisation rate from the flux line lattice which reveals the values of energy gap as $\Delta$(0) = 0.75(3) $meV$ ($\Delta$(0)/k$_{B}$T$_{C}$ = 1.68(6)) for {(HfNb)}$_{0.10}${(MoReRu)}$_{0.90}$ and $\Delta$(0) = 0.95(1) $meV$ ($\Delta$(0)/k$_{B}$T$_{C}$ = 1.96(8)) for {(ZrNb)}$_{0.10}${(MoReRu)}$_{0.90}$. These values are in good agreement with the values obtained from specific heat data. For high $H_{c2}$(0) superconductor, muon spin relaxation rate in the superconducting state ($\sigma_{FLL}$) is related to London penetration depth $\lambda$ via expression \cite{lam1,lam2};
\begin{equation}
\frac{\sigma_{\mathrm{FLL}}^2(T)}{\gamma_{\mu}^2} = \frac{0.00371\Phi_{0}^2}{\lambda^{4}(T)}
\label{eqn13:sigmaH}
\end{equation}
where $\gamma_{\mu}$/2$\pi$ = 135.5 $MHz/T$ is muon gyromagnetic ratio and  $\Phi_{0}$ is the magnetic flux quantum. 
Within the London approximation, the estimated value of $\lambda(0)$ are 495(6) \text{$nm$} and 522(1) \text{$nm$} for {(HfNb)}$_{0.10}${(MoReRu)}$_{0.90}$ and {(ZrNb)}$_{0.10}${(MoReRu)}$_{0.90}$, respectively.

\subsubsection{Discussions} \label{subsubsection:my}

 To supplement the results of the experimental measurements, we have performed calculations of the electronic properties. The electronic heat coefficient $\gamma_{n}$ is directly dependent on effective mass $m^{*}$ and carrier density $n$ of quasi-particle via this expression \cite{Thinkam}. 
\begin{equation}
\gamma_{n} = \left(\frac{\pi}{3}\right)^{2/3}\frac{k_{B}^{2}m^{*}n^{1/3}}{\hbar^{2}}
\label{eqn16:gf}
\end{equation}
where $k_{B}$ is Boltzman constant, using the electronic heat coefficient $\gamma_{n}$ = 3.6(1) and  3.8(1) $mJ$-$mol^{-1}$K$^{-2}$ (determine from normal state heat capacity), and the carrier density $n$ = 11.9(4), 11.4(8)$\times$10$^{28}$m$^{-3}$ (obtain by hall measurement) for ({(HfNb)}$_{0.10}${(MoReRu)}$_{0.90}$ and {(ZrNb)}$_{0.10}${(MoReRu)}$_{0.90}$) respectively. This yield effective mass $m^{*}$ = 4.7(2) $m_{e}$, and 5.3(2) $m_{e}$ respectively for {(HfNb)}$_{0.10}${(MoReRu)}$_{0.90}$ and {(ZrNb)}$_{0.10}${(MoReRu)}$_{0.90}$.
The carrier density $n$ and effective mass $m^{*}$ of quasi-particle are related to Fermi velocity $v_{F}$ by the expression
\begin{equation}
n = \frac{1}{3\pi^{2}}\left(\frac{m^{*}v_{\mathrm{f}}}{\hbar}\right)^{3}
\label{eqn17:n}
\end{equation}
the expression give the Fermi velocity $v_{F}$ = 3.7(2), 3.2(2) $\times {10}^{5} m/s$ for {(HfNb)}$_{0.10}${(MoReRu)}$_{0.90}$ and {(ZrNb)}$_{0.10}${(MoReRu)}$_{0.90}$ respectively. The mean free path $l$ is related to residual resistivity $\rho_{0}$, effective mass $m^{*}$ and Fermi velocity $v_{F}$ of quasi-particle as;
 \begin{equation}
\textit{l} = \frac{3\pi^{2}{\hbar}^{3}}{e^{2}\rho_{0}m^{*2}v_{\mathrm{F}}^{2}}
\label{eqn18:le}
\end{equation}
Using the previous calculated value of effective mass $m^{*}$, the Fermi velocity $v_{F}$ with residual resistivity at transition temperature $\rho_{0}$ =  204(1) $\mu\ohm$-$cm$ and 238(1) $\mu\ohm$-$cm$ (from resistivity measurement), we obtain the electronic mean free path $l$ = 2.6(2) \textup{\AA}, and 2.4(5) \textup{\AA} for {(HfNb)}$_{0.10}${(MoReRu)}$_{0.90}$ and {(ZrNb)}$_{0.10}${(MoReRu)}$_{0.90}$, respectively.  Within the BCS theory, the coherence length $\xi_{0}$ can be expressed in term of Fermi velocity $v_{F}$ and transition temperature $T_{C}$ as 
\begin{equation}
\xi_{0} = \frac{0.18{\hbar}{v_{F}}}{k_{B}T_{C}}
\label{eqn19:xi}
\end{equation}
here $k_{B}$ is Boltzman constant, $v_{F}$ and $T_{C}$ (from magnetization) are the Fermi velocity and transition temperature, by which, we get $\xi_{0}$ = 974(71) \textup{\AA}, 797(64) \textup{\AA} and the ratio of $\xi_{0}$/$l$ $>$ 1 for both {(HfNb)}$_{0.10}${(MoReRu)}$_{0.90}$ and {(ZrNb)}$_{0.10}${(MoReRu)}$_{0.90}$, which is clearly suggesting the signature of dirty limit superconductivity for both the HEA samples. The calculated parameter are listed in Table 1.\\
\begin{figure}
\includegraphics[width=1.0\columnwidth, origin=b]{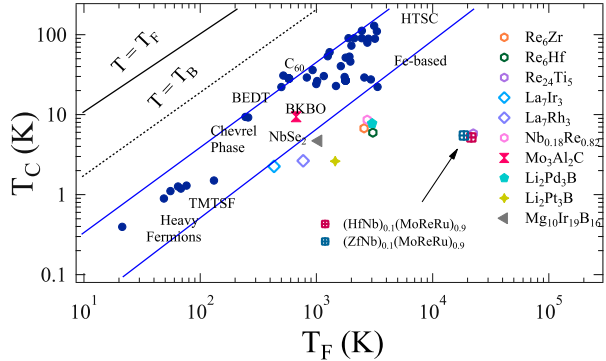}
\caption{A plot of superconducting transition temperature versus Fermi temperature for different superconducting families. Two solid blue lines show the unconventional band of superconductors with other exotic superconductors \cite{elec_prop,R24T5_U,R6Z_U,N0.18R0.82_U} and lies near the unconventional band. {(HfNb)}$_{0.10}${(MoReRu)}$_{0.90}$ and {(ZrNb)}$_{0.10}${(MoReRu)}$_{0.90}$ are shown by red and blue square symbol, which lie far from the other unconventional superconductors.}
\label{Fig8}
\end{figure}
Uemura $et$ $al$. \cite{U1,U2,U3} classified the superconductor in conventional and unconventional on behalf of the ratio of superconducting temperature to the Fermi temperature. If this ratio value falls between 0.01 $\leq$ $T_{c}$/$T_F$ $\leq$ 0.1, then a superconductor is considered as an unconventional one, and heavy fermion superconductor, high $T_{C}$ superconductor, organic superconductor, Fe-based superconductor lie inside this band. To calculating the $T_{F}$ value for both HEA samples, the expression used is as follows:
$k_{B}T_{F}$ = ${\frac{{\hbar}^{2}}{2m^{*}} (3{\pi}^{2}n)^{2/3}}$
where $m^{*}$, $k_{B}$, and $n$ are the effective mass, Boltzmann constant, and carrier density, respectively. The estimated value of $T_{F}$ for {(HfNb)}$_{0.10}${(MoReRu)}$_{0.90}$ and {(ZrNb)}$_{0.10}${(MoReRu)}$_{0.90}$ are 21520(1480) $K$ and 18589(1445) $K$. The $T_{C}/T_{F}$ which are far from the boundary of the unconventional superconductor like the other noncentrosymmetric and unconventional superconductors.

\begin{table}[h!]
\caption{Superconducting and normal state parameters of {(HfNb)}$_{0.10}${(MoReRu)}$_{0.90}$ ({Hf}-HEA) and {(ZrNb)}$_{0.10}${(MoReRu)}$_{0.90}$ ({Zr}-HEA)}
\begin{center}
\begin{tabular}[b]{lccc}\hline\hline
PARAMETERS& UNIT& {Hf}-HEA& {Zr}-HEA\\
\hline
\\[0.5ex]                                  
$T_{C}$& $K$& 5.2(1)& 5.5(1)\\             
$H_{C1}(0)$& $mT$& 2.1(1)& 3.2(3)\\                       
H$_{C2}^{mag}$(0)& $T$& 9.4(1)& 11.5(2)\\
$H_{C2}^{P}(0)$& $T$& 9.56(3)& 10.12(3)\\
$\xi_{GL}$& \text{$nm$}& 5.92(2)& 5.35(3)\\
$\lambda_{GL}$& \text{$nm$}& 609(2)& 487(3)\\
$\lambda(0)_{muon}$& \text{$nm$}& 495(6)& 522(1)\\
$k_{GL}$& &103(1)& 91(1)\\
$\Delta C_{el}/\gamma_{n}T_{C}$&   &1.67(5)& 1.49(4)\\
$\Delta(0)/k_{B}T_{C}$(specific heat)&  &1.85(3)& 1.82(2)\\
$\Delta(0)/k_{B}T_{C}$(muon)&  &1.68(6)& 1.96(8)\\
$m^{*}/m_{e}$& & 4.7(2)& 5.3(2)\\ 
$v_{F}$& $10^{5}ms^{-1}$& 3.7(2)& 3.2(2)\\
$n_s$& 10$^{28}m^{-3}$&11.9(4)& 11.4(8)\\
$\xi_{0}/l_{e}$&   &374(98)& 332(93)\\[0.5ex]
\hline\hline
\end{tabular}
\par\medskip\footnotesize
\end{center}
\end{table}

\section{Conclusion and Summary}

In conclusion, we have performed the first full characterisation of the superconducting properties on the NCS HEAs. In particular, (HfNb)$_{0.10}$(MoReRu)$_{0.90}$ and (ZrNb)$_{0.10}$(MoReRu)$_{0.90}$ by using $\mu$SR, magnetisation, transport and heat capacity measurements. This confirms bulk superconductivity and $H_{C2}$(0) is close to the Pauli limiting field, like other Re-based NCS superconductors. The specific heat and TF-$\mu$SR measurements suggest moderately coupled superconductivity with the isotropic superconducting gap. The ZF-$\mu$SR result indicates time reversal symmetry is preserved in the superconducting ground state for both the HEAs. Comparing the superconducting parameter of the NCS HEA with the binary Re-based compounds (except the preserved time reversal symmetry in the superconducting ground state) are surprisingly similar. The preserved TRS in the Re-based HEA's superconducting ground state despite structural and superconducting properties similarity with Re$_{6}$X series of compounds indicates the complex interplay of disorder and Re in the presence and absence of TRSB in the superconducting ground state. Similarity to binary superconductors and very low heat capacity value (in range of elements) in multicomponent HEA warrant further microscopic studies on more superconducting HEA to understand whether all superconducting HEA alloys show similar behaviour or HEA's with $\alpha $-Mn crystal structure are unique.

\section{Acknowledgments}
Kapil Motla acknowledges the Council of Scientific and Industrial Research (CSIR) Government of India for providing SRF fellowship (Award No. 09/1020(0123)/2017-EMR-I). R.~P.~S.\ acknowledge Science and Engineering Research Board, Government of India for the Core Research Grant CRG/2019/001028. We thank ISIS, STFC, UK for the beamtime to perform the $\mu$SR experiments \cite{DOI}.

\end{document}